\documentclass[letterpaper,aps,pre,twocolumn,showcaps,superscriptaddress,showpacs,amsmath,amssymb]{revtex4}
\usepackage{graphicx}

\newcommand{\dd}{\mathrm{d}}
\newcommand{\pd}[2]{\frac{\partial #1}{\partial #2}}

\newcommand{\mean}[1]{\langle #1 \rangle}
\newcommand{\Int}[1]{\int\dd #1\;}
\newcommand{\IInt}[3]{\int_{#2}^{#3}\dd #1\;}
\newcommand{\Path}[1]{\int[\dd#1]\;}
\newcommand{\unit}[1]{\,\mathrm{#1}}

\newcommand{\gam}{\gamma}
\newcommand{\kap}{\kappa}
\newcommand{\lam}{\lambda}
\newcommand{\vhi}{\varphi}
\newcommand{\sig}{\sigma}
\newcommand{\ra}{\rightarrow}

\newcommand{\x}{\mathbf r}
\newcommand{\q}{\mathbf q}
\newcommand{\tih}{\tilde h}
\newcommand{\kon}{k^\text{on}}
\newcommand{\koff}{k^\text{off}}
\newcommand{\Ha}{\mathcal H}
\newcommand{\Hns}{\Ha_\text{ns}}
\newcommand{\Hs}{\Ha_\text{s}}
\newcommand{\eb}{\epsilon_\text{b}} 
\newcommand{\et}{\epsilon_\text{t}} 
\newcommand{\Eb}{E_\text{b}}
\newcommand{\im}{\text{i}}
\newcommand{\kT}{k_\text{B}T}

\newcommand{\tun}{\tau_\mathrm{un}}

\DeclareMathOperator{\kei}{kei}

\begin{document}


\title{Specific Adhesion of Membranes: Mapping to an Effective Bond Lattice
  Gas}
\author{Thomas Speck}
\affiliation{Department of Chemistry, University of California, Berkeley,
  California 94720, USA}
\affiliation{Chemical Sciences Division, Lawrence Berkeley National
  Laboratory, Berkeley, California 94720, USA}
\author{Ellen Reister}
\author{Udo Seifert}
\affiliation{II. Institut f\"ur Theoretische Physik, Universit\"at Stuttgart,
  D-70550 Stuttgart, Germany.}

\begin{abstract}
  We theoretically consider specific adhesion of a fluctuating membrane to a
  hard substrate via the formation of bonds between receptors attached to the
  substrate and ligands in the membrane. By integrating out the degrees of
  freedom of the membrane shape, we show that in the biologically relevant
  limit specific adhesion is well described by a lattice gas model, where
  lattice sites correspond to bond sites. We derive an explicit expression for
  the effective bond interactions induced by the thermal undulations of the
  membrane. Furthermore, we compare kinetic Monte Carlo simulations for our
  lattice gas model with full dynamic simulations that take into account both
  the shape fluctuations of the membrane and reactions between receptors and
  ligands at bond sites. We demonstrate that an appropriate mapping of the
  height dependent binding and unbinding rates in the full scheme to rates in
  the lattice gas model leads to good agreement.
\end{abstract}

\maketitle


\section{Introduction}

Adhesion of biomembranes to each other is ubiquitous in living organisms since
it is \textit{inter alia} crucial for processes like cell signaling or wound
healing~\cite{beckerle}. While in many cases two membranes may effectively be
attracted to each other due to non-specific interactions, like electrostatic
or van der Waals interactions, adhesion in biological systems is typically
supported by receptors and ligands in the membrane that form bonds upon
contact~\cite{Receptors,Hammer:1996,Bongrand:1999,Wong:2008}. The largest
protein families involved in adhesion are cadherins, integrins, and
selectins~\cite{boal}. In order to gain understanding of the processes of
specific adhesion much experimental effort has been made to develop model
systems using artificially prepared lipid bilayer vesicles with inserted
ligands that are brought into the vicinity of receptors tethered to stiff or
soft, polymer-cushioned,
substrates~\cite{Cuvelier:2004,tana05,Mossman:2007,Smith:2009,Sengupta:2010}.

From a theoretical perspective, membranes are well described as
two-dimensional sheets with a bending rigidity and a rather small effective
surface tension. This model, first introduced by Helfrich~\cite{helf78}, has
been successful in describing morphology and dynamics of free lipid bilayers
and vesicles~\cite{seif97}. Since membranes are often not free but move close
--or adhere-- to other membranes (or, in model systems, to substrates) much
work has been dedicated to the understanding of non-specific interactions
between membranes~\cite{lipo95}. In particular, analogies to wetting theory
and the character of the transition between bound and unbound
membrane~\cite{Lipowsky:1986,lipo94,Sackmann:2002,Komura:2003} have been
explored. In many studies on adhesion receptor-ligand bonds were integrated
out exactly in order to derive an effective interaction potential that depends
on the distance between the membranes or between membrane and
substrate~\cite{lipo96,Zuckerman:1998,Weikl:2000, weik02}. Other studies
employed mean-field approaches that use coarse-grained bond density
fields~\cite{Weikl:2001a,zhan08}. In all of these approaches the discrete
nature of specific adhesion involving receptor-ligand pairs is hidden.  It
only remains visible in more recent work relying on the use of simulations
that typically combine the continuous nature of the membrane with the discrete
nature of the adhesion bonds~\cite{brow08,reis08,weik09}.

\begin{figure}[b]
  \centering
  \includegraphics[scale=.85]{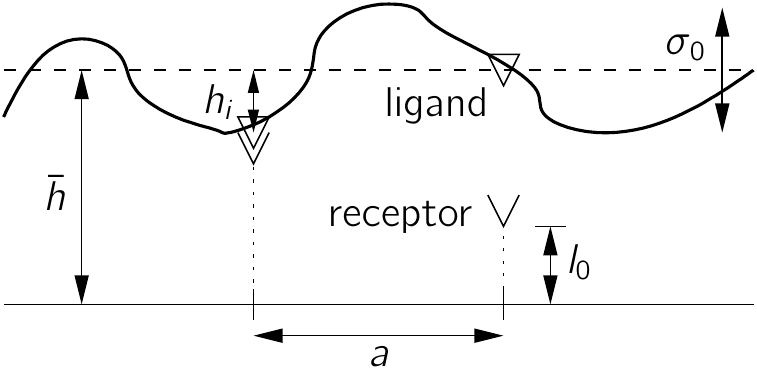}
  \caption{Sketch of the membrane. Besides non-specific interactions between
    substrate and membrane, bonds (with rest length $l_0$) between receptors
    and ligands embedded into the membrane can form. The mean height of the
    membrane with respect to the substrate is $\bar h$. Thermal fluctuations
    of the membrane are of size $\sig_0$. Above bond $i$, the deviation of the
    membrane height profile from the mean height is $\sig_0h_i$.}
  \label{fig:mem}
\end{figure}

In this work, we follow the opposite strategy compared to previous theoretical
studies by integrating out the membrane shape in order to describe adhesion
purely through receptor-ligand bonds. Starting from a mixed
continuous-discrete model we show that in the regime of confined and rather
stiff membranes, which is relevant for cells, it maps onto a lattice gas,
where lattice sites correspond to receptor-ligand pairs. Membrane properties
enter the effective interactions between bonds. The benefits from such an
approach are two-fold: firstly, we can apply the vast knowledge and methods
developed for lattice gases to the adhesion behavior of membranes, and,
secondly, it allows for the efficient numerical simulation by only taking into
account the discrete bonds. We demonstrate in which regime such a pure lattice
gas description is valid and explicitly calculate the effective bond
interactions. By constructing realistic reaction rates taking into account the
distance between the membrane and the substrate, we are capable of
qualitatively capturing the dynamics of adhesion within the lattice gas
framework. We support this claim by comparing simulation results for the full
Hamiltonian with lattice gas simulations employing the kinetic Monte Carlo
method.


\section{Model and mapping}

We consider a fluctuating bio-membrane in a fluid at temperature $T$ above a
substrate. The membrane is described in the Monge representation by the height
profile $h(\x)$ with $\x=(x,y)$. The projected area is $A$ and we employ
periodic boundary conditions. In addition, $N$ receptors at positions
$\{\x_i\}$ are embedded in the substrate that can bind to the corresponding
ligands in the membrane, see Fig.~\ref{fig:mem}.

Our starting point is the total Hamiltonian $\Ha=\Ha_0+\Hns+\Hs$ composed of
three terms~\cite{weik02,reis08}. The first term is the Helfrich
energy
\begin{equation}
  \label{eq:H:0}
  \Ha_0[h(\x)] = \IInt{\x}{A}{} \frac{\kap}{2}[\nabla^2h(\x)]^2
\end{equation}
with bending rigidity $\kap$ governing the thermal fluctuations of the
membrane. Non-specific interactions between the substrate and the membrane are
due to, e.g., surface charges, van der Waals forces, or steric repulsion. They
occur over ranges where individual molecules are not resolved. These
non-specific interactions are modeled through the simple quadratic potential
\begin{equation}
  \label{eq:H:ns}
  \Hns[h(\x)] = \IInt{\x}{A}{} \frac{\gam}{2} [h(\x)-h_0]^2
\end{equation}
with strength $\gam$ resulting from a Taylor expansion around the minimum at
height $h_0$. The magnitude of the membrane height fluctuations in the absence
of bonds is
\begin{equation}
  \label{eq:14}
  \begin{split}
    \sig_0^2 &\equiv \mean{[h(0)-\bar h]^2} \\
    &= \frac{1}{2\pi\beta}\IInt{q}{0}{\infty}\frac{q}{\kap q^4+\gam}
    = \frac{1}{8\beta\sqrt{\gam\kap}}
  \end{split}
\end{equation}
using $\mean{|\tih_\q|^2}=[\beta A(\kap q^4+\gam)]^{-1}$ with
$\beta\equiv1/\kT$ (see Eq.~\eqref{eq:fourier} for the definition of the
Fourier coefficients). Here, $\bar h$ is the mean height of the membrane.

The focus of this work lies on specific substrate-membrane interactions
through ligand-receptor pairs contributing with
\begin{equation}
  \label{eq:H:s}
  \Hs[h(\x);\{b_i\}] = \sum_{i=1}^N b_i \left\{
    \frac{k}{2}(\bar h+\sig_0h_i-l_0)^2 - \eb \right\}
\end{equation}
to the total Hamiltonian. This energy is determined by two competing terms,
the specific binding energy $\eb$ and the energy stored in a stretched bond
with effective spring constant $k$. The $N$ binary variables $b_i$ are 1 for a
closed bond and 0 for an open bond. The bonds have rest length $l_0$, and
$h(\x_i)\equiv\bar h+\sig_0h_i$ is the distance of the membrane ligand at
$\x_i$ from the substrate. The dimensionless $h_i$ are of order unity. It will
be advantageous to define two more quantities: the (average) area per bond
\begin{equation*}
  a^2 \equiv A/N
\end{equation*}
and the fraction
\begin{equation*}
  \phi \equiv \frac{1}{N} \sum_{i=1}^N b_i
\end{equation*}
of closed bonds.

The partition sum reads
\begin{equation}
  \label{eq:Z}
  Z \equiv \sum_{\{b_i\}} \Path{h(\x)} e^{-\beta\Ha[h(\x);\{b_i\}]}.
\end{equation}
Our goal is to derive an effective lattice gas model by integrating out the
height field $h(\x)$. The integration over the membrane fluctuations under the
constraint of given heights $h_i$ is carried out in appendix~\ref{sec:fluct},
\begin{multline}
  \label{eq:Z:2}
  Z \simeq \sum_{\{b_i\}} \Int{h_1\cdots\dd h_N}
  e^{-\frac{1}{2}\sum_{ij}(m^{-1})_{ij}h_ih_j - \beta \Eb},
\end{multline}
leading to a coupling of the height variables $h_i$ with coupling matrix
\begin{equation}
  \label{eq:m}
  m_{ij} \equiv \frac{2}{\beta A\sig_0^2}
  \sum_{\q\neq0} \frac{\cos\q\cdot(\x_i-\x_j)}{\kappa q^4+\gam},
\end{equation}
where the sum runs over independent wave vectors only. We rewrite the explicit
expression for $\Eb$ from Eq.~(\ref{eq:app:Eb}) in terms of three
dimensionless parameters $\chi$, $\theta$, and $\lam$:
\begin{multline}
  \label{eq:Eb}
  \beta\Eb = -\frac{1}{2}\frac{[\chi\sum_ib_ih_i-N\theta\lambda]^2}{N\theta+N
    \phi \chi} \\
  + \frac{1}{2}\theta\lambda^2 + \sum_{i=1}^N b_i \left(\frac{1}{2}\chi
    h_i^2 - \beta\eb\right).
\end{multline}
Here,
\begin{equation}
  \label{eq:chi}
  \chi \equiv \beta k\sig_0^2
\end{equation}
is an effective cooperativity parameter between bonds discussed further below,
while
\begin{equation}
  \theta \equiv \beta \gamma a^2\sig_0^2
\end{equation}
is the average energy contribution from the non-specific potential per
bond. Finally, $\lambda\equiv(h_0-l_0)/\sig_0$ is the ratio of the distance
between the minimum of the non-specific potential and the rest length of the
tether, and the amplitude of height fluctuations.

For now we assume that $\chi$ is small. Below and in the next section we will
demonstrate that this assumption indeed corresponds to the range of parameters
we are interested in. Hence, Taylor-expanding Eq.~(\ref{eq:Eb}) up to first
order in $\chi$ we obtain
\begin{equation*}
  \beta\Eb \approx \chi\lam \sum_{i=1}^N b_ih_i
  + \sum_{i=1}^N b_i\left[\frac{1}{2}\chi h_i^2 - \beta(\eb-\et) \right]
\end{equation*}
with the typical tether energy
\begin{equation}
  \label{eq:eps}
  \et \equiv \frac{k}{2}(h_0-l_0)^2 = \frac{1}{2}\chi\lam^2\kT.
\end{equation}
Within this approximation $\Eb$ becomes independent of $\theta$. Instead of
$\lam$ we will take $\et$ as the independent variable in the following,
leaving us with $\et$, $\eb$, and $\chi$. We now perform the Gaussian
integration in Eq.~(\ref{eq:Z:2}). To this end, we have to invert the matrix
$(m^{-1})_{ij}+\chi b_i\delta_{ij}$, leading to $m_{ij}+\mathcal
O(\chi)$. Hence, to first order in $\chi$ we finally arrive at
\begin{equation}
  \label{eq:Z:lq}
  Z \simeq \sum_{\{b_i\}} \exp\left\{ \beta\sum_{i\neq j}\nu_{ij}b_ib_j +
    \beta\mu\sum_{i=1}^Nb_i\right\}
\end{equation}
which is isomorphic to the lattice gas model of the liquid-gas transition with
a binary density $b_i$~\cite{chandler}. The sum runs over all bonds where the
diagonal term is included in the lattice gas chemical potential
\begin{equation}
  \label{eq:mu}
  \mu \equiv \eb - \et(1-\chi).
\end{equation}
The effective interaction energy between bonds is given through
\begin{equation}
  \label{eq:nu}
  \nu_{ij} \equiv \et\chi m(|\x_i-\x_j|), \quad
  m(r) = -\frac{4}{\pi}\kei_0(r/\xi)
\end{equation}
with length scale
\begin{equation}
  \label{eq:xi}
  \xi \equiv (\kap/\gam)^{1/4} = \sqrt{8\beta\kap}\,\sig_0 =
  \sqrt{8\chi\kap/k}.
\end{equation}
For the derivation and form of this effective interaction, see
appendix~\ref{sec:eff} and Fig.~\ref{fig:pot}.

\begin{figure}[t]
  \centering
  \includegraphics[width=\linewidth]{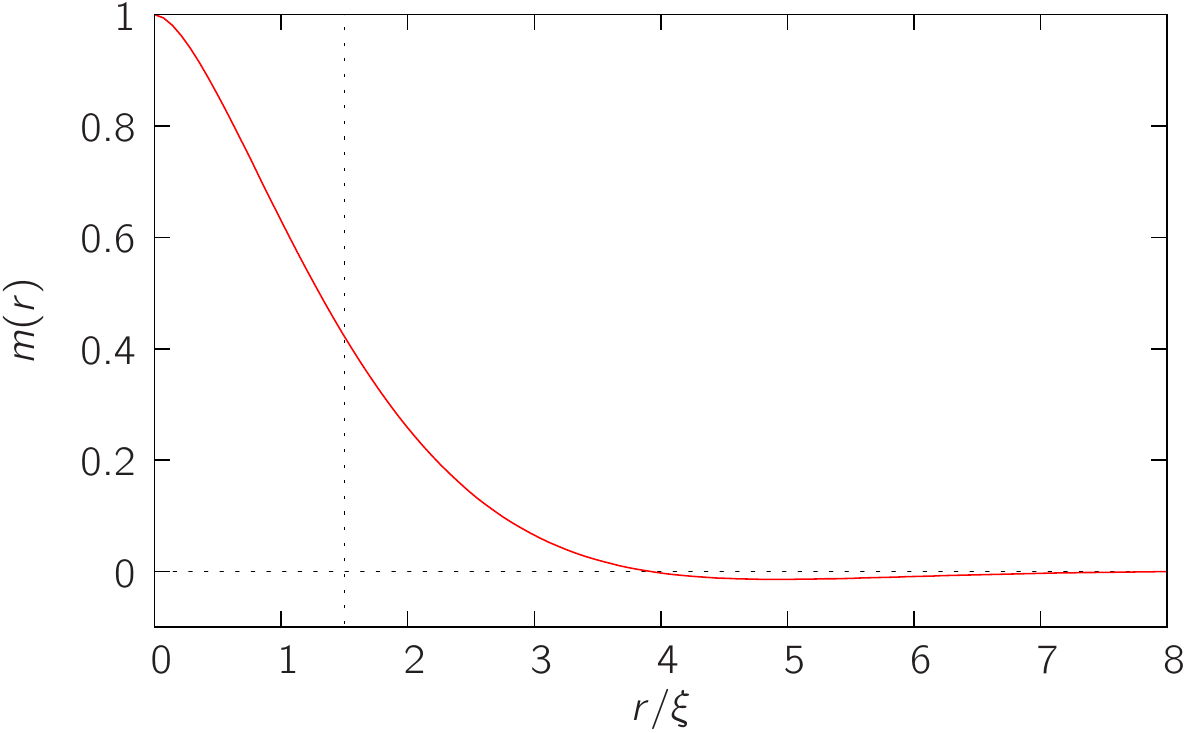}
  \caption{Decay of the effective membrane-induced interactions $m(r)$ between
    bonds from Eq.~(\ref{eq:nu}). The dotted vertical line marks the bond
    distance $a$ used for the nearest-neighbor approximation.}
  \label{fig:pot}
\end{figure}


This mapping from the original full model to an effective bond lattice gas is
the central result of this paper valid for any geometry of bonds. The small
quantity $\chi$ defined in Eq.~(\ref{eq:chi}) determines the \textit{local
  cooperative} interactions of bonds. The physical picture is that a closed
bond pulls down the membrane, assisting neighboring bonds to form. We have
assumed the local effect of this behavior to be small; either through weak
links (small $k$) such that the deforming force on the membrane is small, or
by a stiff, confined membrane (small $\sig_0$) which is rather pulled down as
a whole instead of being locally deformed. The membrane mediated interactions
between bonds then decay on the length scale $\xi$, and an effective
description in terms of the bonds alone becomes feasible.

\section{Square lattice: nearest-neighbor approximation}

For a perfectly flat membrane with $\sig_0=0$ (implying $\chi=0$) there is no
coupling between bonds, i.e., the bonds are independent. The partition sum is
then calculated easily as
\begin{equation}
  \label{eq:2}
  Z = \prod_{i=1}^N \sum_{b_i} e^{\beta\mu b_i} 
  = \left[1+e^{\beta(\eb-\et)}\right]^N.
\end{equation}
The mean bond density
\begin{equation}
  \label{eq:6}
  \mean{\phi} = \frac{1}{N}\pd{\ln Z}{(\beta\mu)}
  = \frac{e^{\beta(\eb-\et)}}{1+e^{\beta(\eb-\et)}}
\end{equation}
shows a continuous crossover between a bound (for $\eb\gg\et$) and an unbound
(for $\eb\ll\et$) state with $\mean{\phi}=1/2$ for $\eb=\et$.

\begin{figure}[t]
  \centering
  \includegraphics[scale=.8]{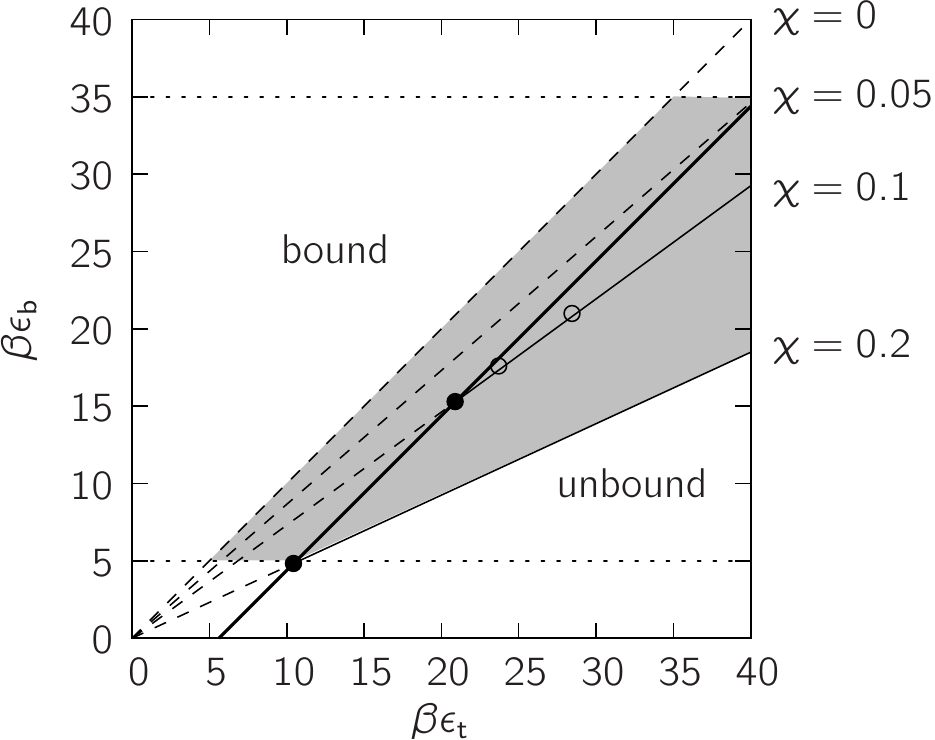}
  \caption{Set of equilibrium phase diagrams for bonds on a square lattice
    with nearest-neighbor interactions ($m_1\simeq0.42$). Critical points
    (filled circles) fall on a line (thick line). The shaded region indicates
    the accessible range of coexistence (solid) and crossover (dashed) lines,
    and the relevant specific binding energies. In Sec.~\ref{sec:dynamics} we
    investigate the dynamics of two state points marked by the open circles
    for $\chi=0.1$.}
  \label{fig:phase}
\end{figure}

For a more quantitative analysis of a fluctuating membrane with $\sig_0>0$ we
exploit the fact that the effective interactions decay exponentially fast, see
Fig.~\ref{fig:pot}. Therefore, we can approximate the bond interactions by
taking into account only nearest-neighbors. Moreover, we assume the
ligand-receptor pairs to be arranged on a square lattice with spacing $a$. For
an upper bound of the validity of this regime, we choose a lattice spacing
$a=\tfrac{3}{2}\xi$ with $m_1\equiv m(a)\simeq0.42$ and interaction energy
$\nu=\et\chi m_1$. Exploiting the analogy with the lattice gas we can
immediately specify the equilibrium phase diagram for independent energies
$\eb$ and $\et$ parametrized by $\chi$, see Fig.~\ref{fig:phase}. Using that
at phase coexistence the chemical potential is $\mu_\text{co}=-4\nu$ and that
the critical points obey $\beta\nu^\ast\simeq 2/2.269$~\cite{chandler}, the
line of critical points is determined as
\begin{equation}
  \label{eq:crit}
  \eb^\ast = \et^\ast - \frac{\nu^\ast}{m_1}(1+4m_1)
\end{equation}
with slope one, where the offset depends on $m_1$. Specifying $\chi$ selects a
single critical point. At every critical point, a coexistence line
$\eb(\et)=z\et$ with slope
\begin{equation}
  \label{eq:z}
  z \equiv 1-\chi(1+4m_1)
\end{equation}
for $\et>\et^\ast$ terminates. The straight continuation of the coexistence
line below the critical point $\et<\et^\ast$ marks the continuous crossover
between bound and unbound membrane where $\mean{\phi}=1/2$.

We restrict our discussion to typical values for the specific binding energy
$\eb$ ranging from $5\kT$ to $35\kT$~\cite{webe75,boal} as indicated in
Fig.~\ref{fig:phase}. The critical point corresponding to the upper limit
implies $\chi_1\simeq0.05$. Hence, in the range $0\leqslant\chi<\chi_1$ for
$\eb\leqslant35\kT$ the qualitative picture for $\chi=0$ persists and a
continuous crossover is observable. For $\chi>\chi_1$ the critical point
enters the relevant parameter region and the first order transition becomes
accessible. Increasing $\chi$ further, the critical point moves down the line
given by Eq.~\eqref{eq:crit} to smaller specific binding energies, eventually
reaching the lower limit $\eb^\ast\simeq5\kT$ for $\chi\simeq0.2$. Hence, the
shaded area in Fig.~\ref{fig:phase} marks the set of coexistence and crossover
lines of the effective lattice gas that is compatible with our initial
assumption of a small cooperativity parameter $\chi$. A genuine coexistence
between closed bonds and open bonds requires both non-negligible membrane
fluctuations and rather large specific binding energies $\eb$ and tether
energies $\et$. For either weak fluctuations or weak binding energies only a
continuous crossover between these two states is obtained.

Pushing the analogy with the lattice gas further, in the first order
transition regime we estimate the free energy barrier an initially unbound
membrane needs to overcome in order to bind to the substrate. From the
partition sum~\eqref{eq:Z:lq} we extract the free energy
\begin{equation*}
  G(\{b_i\}) = \nu\sum_{\mean{ij}}(b_i-b_j)^2 -
  (\eb-z\et)\sum_{i=1}^N b_i,
\end{equation*}
where the sum in the first term runs over all bond pairs. It counts the number
of 'broken' links (two neighboring bonds in different states) adding up to the
interface length between bound and unbound domains. Assuming a circular domain
of size $n$ in the spirit of Becker-D\"oring~\cite{beck35}, the change of free
energy associated with this domain is
\begin{equation*}
  \Delta G(n) \simeq 2\nu\sqrt{\pi n} - (\eb-z\et)n.
\end{equation*}
The critical size of the nucleus and the barrier height are
\begin{equation*}
  n^\ddagger \simeq \frac{\pi\nu^2}{(\eb-z\et)^2}, \qquad
  \Delta G^\ddagger \simeq \frac{\pi\nu^2}{\eb-z\et},
\end{equation*}
respectively. The nucleation time then grows exponentially with $\beta\Delta
G^\ddagger$.


\begin{figure*}[t]
  \centering
  \includegraphics[width=\textwidth]{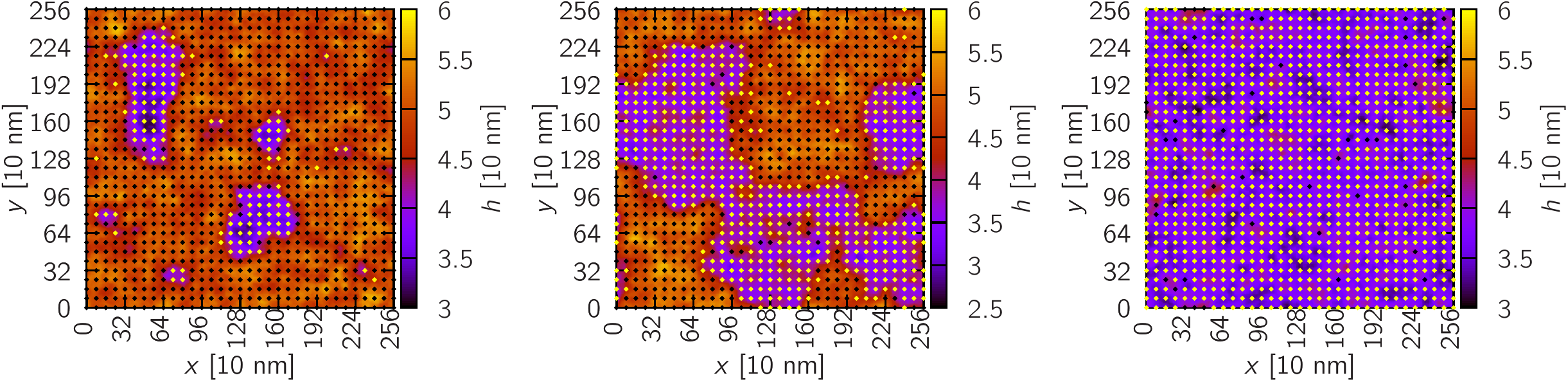}
  \caption{Membrane height profile and bonds for a single run of an initially
    unbound membrane: when the critical nucleus has formed (left), when the
    bond density has reached $\phi\simeq1/2$ (center), and in equilibrium
    (right). A closed bond is drawn light, an open bond is dark. Parameters
    as in Fig.~\ref{fig:dyn}b).}
  \label{fig:snap}
\end{figure*}

\section{Dynamics}
\label{sec:dynamics}

Going beyond static properties, in this section we compare the effective bond
lattice gas with the full model for the adhesion dynamics close to
coexistence. Integrating out the membrane degrees of freedom, one expects the
description of bond dynamics through a Markovian stochastic process without
memory to be a good approximation in the limit where the formation of bonds is
slow compared to the time scale of membrane undulations.

\subsection{Binding and unbinding rates}

We start with the implementation of dynamics in the full model. In Fourier
space, the membrane modes obey the Langevin equation
\begin{equation}
  \label{eq:5}
  \partial_t\tih_\q = -\Lambda_\q\pd{\Ha}{\tih^\ast_\q} + \zeta_\q,
\end{equation}
where $\Lambda_\q$ are the Onsager coefficients that take into account the
fluid surrounding the membrane~\cite{seif97} and $\zeta_\q$ is Gaussian white
noise obeying the fluctuation-dissipation theorem. Further details of how the
dynamics of the membrane fluctuations is implemented in the full simulation
scheme are discussed in Ref.~\cite{reis07,reis10}.

After a membrane step the heights $h(\x_i)$ are extracted via discrete Fourier
transformation and the bonds are updated. The membrane exerts a force $f_i$ on
a closed bond that is given by the effective spring constant $k$ times the
separation. The unbinding rates for a bond to rupture are then modeled as the
usual Bell rates~\cite{bell78},
\begin{equation}
  \label{eq:11}
  \koff_i = w_0 e^{\beta f_i \sig_0}, \qquad
  f_i = k[h(\x_i)-l_0]
\end{equation}
with bare dissociation rate $w_0$ and effective size of the binding potential
set to $\sig_0$. The binding rates follow through the detailed balance
condition as,
\begin{equation}
  \label{eq:4}
  \kon_i = \koff_i e^{-\frac{1}{2}\beta k[h(\x_i)-l_0]^2+\beta\eb}.
\end{equation}
These rates are, therefore, independent of the state of other
bonds. Cooperative dynamics between bonds, e.g., the assistance of bond
formation, are completely due to interactions mediated by the membrane. In
Fig.~\ref{fig:snap}, a representative example of the membrane nucleation
dynamics of the full model is shown.

For constructing the dynamics in the effective lattice gas, we do not have the
explicit height information $h(\x_i)$ of the membrane anymore. The most naive
approximation would be to assume a constant unbinding rate. Somewhat more
realistically, we can construct an approximation of the membrane force
$f_i=kl_i(\{b_i\})$ based on the configuration of neighboring bonds. As a
first implementation we will employ a simple linear interpolation
$l_i=l(\phi)$ of the bond separation depending only on the bond density $\phi$
at a given time. The \textit{typical} height of the unbound membrane is
$h_0$. Force balance between tether force and non-specific interactions for a
single closed bond implies
\begin{equation*}
  a^2\gam(h_1-h_0) + k(h_1-l_0) = 0
\end{equation*}
for the typical height $h_1$ of the bound membrane. Linear interpolation then
leads to the expression
\begin{equation*}
  l(\phi) = h_0(1-\phi) + h_1\phi - l_0
  = (h_0-l_0)\left[1-\frac{k\phi}{a^2\gam+k}\right]
\end{equation*}
for the bond separation. Hence, for the unbinding rate we obtain
\begin{equation}
  \label{eq:3}
  k^\text{LG,off}(\phi) = w^\text{LG}_0
  \exp\left\{\sqrt\frac{2\beta\nu}{m_1}\left(1-\frac{32\chi}{9+32\chi}\phi
    \right)\right\}
\end{equation}
in the reduced quantities. The rate $w^\text{LG}_0$ determines the time scale
of the lattice gas. Since the membrane undulations slow down the bond
formation in the full model, this rate will be smaller than $w_0$. The binding
rate again follows through detailed balance
\begin{equation}
  \label{eq:1}
  k^\text{LG,on}_i = k^\text{LG,off} e^{-\beta\nu\sum_{j(i)} b_j-\beta\mu}
\end{equation}
and thus depends on the state of the neighboring bonds.

\begin{figure*}[t]
  \centering
  \includegraphics[width=\textwidth]{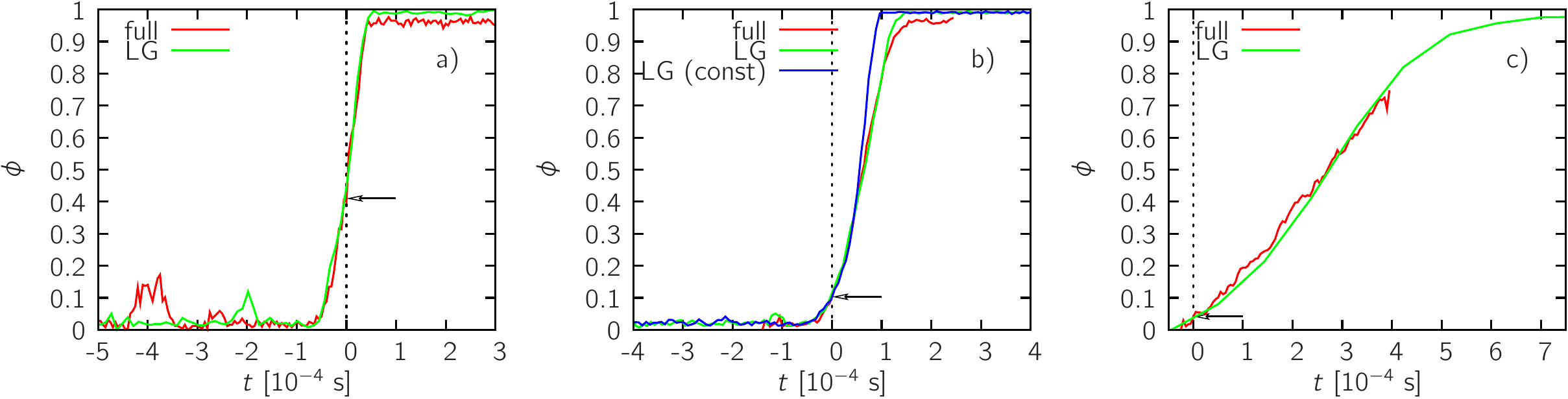}
  \caption{Dynamics of the lattice gas (LG) compared to the full model for
    $\chi\simeq 0.1$ (parameters specified in the main text). Shown is the
    density $\phi(t)$ for an initially unbound membrane in the nucleation
    regime close to the coexistence line with $\beta\et\simeq28.4$,
    $\beta\eb\simeq21.0$ on a) 128$\times$128 and b) a 256$\times$256 grid. c)
    Closer to the critical point with $\beta\et\simeq23.7$,
    $\beta\eb\simeq17.6$ on a 256$\times$256 grid. The arrows mark the
    nucleation barrier $\phi^\ddagger=n^\ddagger/N$.}
  \label{fig:dyn}
\end{figure*}

\subsection{Numerics}

For the numerical simulations we set the bending rigidity $\beta\kap=80$ and
the non-specific strength $\beta\gam=10^{-5}\unit{nm^{-4}}$. The bond distance
is $a=80\unit{nm}$ with bond strength $\beta
k=2.25\times10^{-2}\unit{nm^{-2}}$. These values lead to height fluctuations
$\sig_0\simeq2.1\unit{nm}$, decay length $\xi\simeq53.3\unit{nm}$, and slope
$z\simeq0.73$ of the coexistence line. The cooperativity parameter is
$\chi\simeq0.1$.

In Fig.~\ref{fig:dyn}, binding dynamics in the nucleation regime of an
initially unbound membrane are compared to the bond lattice gas employing
kinetic Monte Carlo moves for two different state points (indicated in
Fig.~\ref{fig:phase}). For Fig.~\ref{fig:dyn}a) and b), we choose
$\beta\et\simeq28.4$ (corresponding to $\beta\nu=1.2$,
$h_0-l_0\simeq50.3\unit{nm}$) and $\beta\eb=z\beta\et+0.2\simeq21.0$. For
Fig.~\ref{fig:dyn}c), we move closer towards the critical point with
$\beta\et\simeq23.7$ ($\beta\nu=1$, $h_0-l_0\simeq45.9\unit{nm}$) and
$\beta\eb=z\beta\et+0.27\simeq17.6$. The plotted mean densities $\phi(t)$ are
averaged over multiple runs, where single runs are shifted such that at $t=0$
the critical nucleus has formed. Fig.~\ref{fig:dyn}b) shows a system of
256$\times$256 receptor-ligand pairs compared to a 128$\times$128 grid in
Fig.~\ref{fig:dyn}a). The curves are qualitatively the same for both system
sizes, however, the waiting time (not shown) for the critical nucleus to form
is much smaller in the bigger system, which is to be expected. The
fluctuations of the bound membrane are larger in the full model compared to
the lattice gas. For comparison, in Fig.~\ref{fig:dyn}b) we show the lattice
gas dynamics for constant unbinding rate $k^\text{LG,off}(0)$. While it
follows the curve of the full model for $\phi<0.5$, it misses to capture the
slowing down of the dynamics for higher bond density. The mean-field rates
$k^\text{LG,off}(\phi)$ perform much better but still for $\phi>0.85$ the
lattice gas dynamics for the almost bound membrane becomes faster compared to
the full model. Moving towards the critical point, dynamics of bond formation
slows down as is clearly visible from Fig.~\ref{fig:dyn}c).

The bare dissociation rate in the full model is set to
$w_0=5\times10^7\unit{s^{-1}}$. This rate is close to the inverse relaxation
time of membrane undulations on the length scale of the bond distance for the
unbound membrane, $\tun\sim\eta a^3/\kap\sim10^{-7}\unit{s}$ with the
viscosity of water $\eta\simeq10^{-3}\unit{Pa\cdot s}$. Such a correspondence
implies that we test a dynamic regime where there is no large time scale
separation between membrane undulations and bond formation. Nevertheless, the
binding dynamics in Fig.~\ref{fig:dyn} show a very good agreement between full
model and lattice gas. Realistic dissociation rates are typically somewhat
smaller and, therefore, in a regime where the lattice gas will perform even
better. The rates $w^\text{LG}_0$ have been extracted from scaling the time
evolution to fit the full model and have been determined as:
$w^\text{LG}_0/w_0\simeq1.8\times10^{-3}$, $2.8\times10^{-3}$,
$2.1\times10^{-4}$ for Fig.~\ref{fig:dyn}a)-c), respectively.


\section{Conclusions}

Beginning with a Hamiltonian that incorporates the essentials of specific
adhesion to a flat substrate, i.e., bending energy of the membrane, a
non-specific adhesion potential, and the energy contribution of
receptor-ligand pairs, we integrate out the membrane shape in the partition
function of the system. In the limit of small membrane roughness and/or weak
stiffnesses of the tethers attaching the receptors to the substrate, our
approach leads to an effective free energy that is isomorphic to that of a
lattice gas model used to describe liquid-gas transitions. We are able to give
an explicit expression for the effective interaction potential between
bonds. When the bond sites are far enough apart it is sufficient to consider
only nearest-neighbor interactions. For bonds on a square lattice the
equilibrium phase behavior follows immediately. The analogy to lattice gas
models also extends to nucleation theory, yielding expressions for the
critical nucleus size of an adhered membrane patch and the energy barrier that
needs to be crossed.

We further show that the lattice gas model may also be utilized to study the
dynamics of the adhesion process. To this end, we compare simulations of the
full Hamiltonian with a kinetic Monte Carlo scheme for the lattice gas. The
comparison of results of the computationally costly simulations of the full
model with the fast lattice gas simulations reveals a very good qualitative
agreement in the increase of the average bond number as a function of
time. We, therefore, expect our model to be useful in more complex studies of,
e.g., moving ligands and the influence of defects on membrane adhesion.


TS gratefully acknowledges financial support by the Alexander von Humboldt
foundation and the Helios Solar Energy Research Center which is supported by
the Director, Office of Science, Office of Basic Energy Sciences of the
U.S. Department of Energy under Contract No.~DE-AC02-05CH11231. US
acknowledges financial support by the DFG under grant SE1119/2. US and ER
thank A.-S. Smith for many stimulating discussions and an ongoing fruitful
collaboration.


\appendix
\section{Membrane fluctuations}
\label{sec:fluct}

We expand the height profile $h(\x)$ into Fourier modes,
\begin{equation}
  \label{eq:fourier}
  h(\x) = \sum_\q \tih_\q e^{-\im\q\cdot\x},
\end{equation}
where $\tih^\ast_\q=\tih_{-\q}$. For a square system of size $L$, the
accessible wave vectors are $\q=(2\pi/L)(n_x,n_y)$ with integer $n_x$, $n_y$,
and $q\equiv|\q|$. The Helfrich energy~(\ref{eq:H:0}) becomes
\begin{equation*}
  \Ha_0 = \frac{\kappa A}{2}\sum_{\q\neq 0} q^4 |\tih_\q|^2.
\end{equation*}
In the non-specific interactions~(\ref{eq:H:ns})
\begin{equation*}
  \Hns = \frac{\gam A}{2}\left\{ (\bar h-h_0)^2 +
    \sum_{\q\neq 0}|\tih_\q|^2 \right\}
\end{equation*}
we split off the zero-mode $\bar h\equiv\tih_0$ contribution.

We want to integrate out the fluctuations of the membrane under the constraint
that at $\x_i$ the height is $\bar h+\sig_0h_i$. This constraint is expressed
as
\begin{multline*}
  \prod_{i=1}^N \delta(h(\x_i)-\sig_0h_i) = \frac{1}{(2\pi)^N}
  \Int{\lam_1\cdots\dd\lam_N} \\ \times
  \exp\left\{\im\sum_{i=1}^N \lam_i[h(\x_i)-\bar h-\sig_0h_i] \right\}.
\end{multline*}
In terms of Fourier modes we have
\begin{equation*}
  h(\x_i) = \bar h + 
  \sum_{\q\neq0}{'} \left[\tih_\q e^{-\im\q\cdot\x_i}+\tih_{-\q}e^{\im\q\cdot\x_i}\right],
\end{equation*}
where the sum $\sum'_\q$ runs over independent modes $\q$ (for a $\q$ exclude
$-\q$ from the sum). Splitting the coefficient $\tih_\q=\tih'_\q+\im\tih''_\q$
into real and imaginary part, the integral over all modes (except the
zero-mode) becomes a product of two independent integrals. These are simple
Gaussians and read
\begin{multline*}
  Z' = \int\prod_{\q\neq0}\dd\tih'_\q \\ \times
  \exp\left\{ -\sum_{\q\neq0}{'} \beta A(\kappa q^4+\gam)
    (\tih'_\q)^2 + \im \sum_{\q\neq0}{'} c'_\q\tih'_\q \right\}
\end{multline*}
for the integration over real parts. Analogously, $Z''$ is obtained through
replacing $\tih'_\q$ by $\tih''_\q$ and $c'_\q$ by $c''_\q$. The coefficients
are
\begin{equation*}
  c'_\q = 2\sum_{i=1}^N \lam_i \cos(\q\cdot\x_i), \quad
  c''_\q = 2\sum_{i=1}^N \lam_i \sin(\q\cdot\x_i).
\end{equation*}
The result is
\begin{equation*}
 Z'Z'' \sim 
  \exp\left\{ -\frac{1}{2} \sig_0^2 \sum_{ij} m_{ij}\lam_i\lam_j \right\}
\end{equation*}
with coupling matrix
\begin{equation*}
  m_{ij} \equiv \frac{2}{\beta A\sig_0^2}
  \sum_{\q\neq0}{'} \frac{\cos\q\cdot(\x_i-\x_j)}{\kappa q^4+\gam}.
\end{equation*}
We notationally suppress the non-exponential prefactors. The integral over the
$\lam_i$ implementing the constraints again is a simple multi-dimensional
Gaussian integral,
\begin{multline*}
  \frac{1}{(2\pi)^N} \Int{\lam_1\cdots\dd\lam_N} \\ \times
  \exp\left\{ -\frac{1}{2}\sig_0^2\sum_{ij} m_{ij}\lam_i\lam_j 
    -\im\sum_i\sig_0h_i\lam_i \right\} \\
  = [(2\pi)^N\det m]^{-1/2} \exp\left\{ -\frac{1}{2}\sum_{ij} 
    (m^{-1})_{ij}h_ih_j \right\}.
\end{multline*}
The remaining energy is a function of the mean height,
\begin{equation*}
  E(\bar h) \equiv \frac{\gam A}{2}(\bar h-\delta h_0)^2
  + \sum_{i=1}^N b_i\left\{ \frac{k}{2}(\bar h+\sig_0h_i)^2 - \eb \right\},
\end{equation*}
where we have shifted $\bar h\ra\bar h+l_0$ for convenience and defined
$\delta h_0\equiv h_0-l_0$. We finally integrate over the mean height
\begin{equation*}
  \Int{\bar h} e^{-\beta E(\bar h)} \simeq e^{-\beta \Eb}
\end{equation*}
with
\begin{multline}
  \label{eq:app:Eb}
  \Eb = -\frac{1}{2}\frac{(k\sig_0\sum_ib_ih_i-\gam A\delta h_0)^2}{\gam
    A+kN\phi} \\
  + \frac{\gam A}{2}(\delta h_0)^2  + \sum_ib_i\left(\frac{k}{2}\sig_0^2h_i^2
    -\eb \right).
\end{multline}

\section{Effective interaction}
\label{sec:eff}

To obtain a more useful expression for the bond interactions, we approximate
the sum over modes in Eq.~(\ref{eq:m}) by a two-dimensional integral,
\begin{align*}
  m(r) &= \frac{\xi^4}{2\pi^2\beta\kap\sig_0^2}
  \IInt{q}{0}{\infty}\IInt{\vhi}{0}{\pi}
  \frac{q\cos[qr\cos\vhi]}{1+(\xi q)^4} \\
  &= \frac{\xi^4}{2\pi\beta\kap\sig_0^2} \IInt{q}{0}{\infty}
  \frac{qJ_0(qr)}{1+(\xi q)^4},
\end{align*}
where $\xi\equiv(\kap/\gam)^{1/4}$ and $J_0(x)$ is the zero-order Bessel
function of the first kind. Performing the integral~\cite{gradshteyn} leads to
the effective interaction
\begin{equation*}
  m(r) = -\frac{1}{2\pi\beta\kap(\sig_0/\xi)^2} \kei_0(r/\xi)
  = -\frac{4}{\pi} \kei_0(r/\xi).
\end{equation*}
Since $\kei_0(0)=-\pi/4$ we have $m(0)=1$. Here, $\kei_0(x)$ is a Kelvin
function defined as
\begin{equation}
  \label{eq:kei}
  \kei_0(x) \equiv \mathrm{Im} K_0(x e^{\im3\pi/4}),
\end{equation}
where $K_0(z)$ is the zero-order modified Bessel function of the second kind.


\end{document}